\documentclass{article}

   \PassOptionsToPackage{square,sort,comma,numbers}{natbib}

 \usepackage[final]{neurips_2020}

\usepackage[utf8]{inputenc} 
\usepackage[T1]{fontenc}  
\usepackage{hyperref}    
\usepackage{url}      
\usepackage{booktabs}    
\usepackage{amsfonts}    
\usepackage{nicefrac}    
\usepackage{microtype}   

\usepackage{enumitem}
\usepackage{multirow}
\usepackage{graphicx}
\usepackage{comment}
\usepackage{float}

\title{Is Transfer Learning Necessary for Protein Landscape Prediction?}

\author{%
 Amir Shanehsazzadeh \\
 Harvard University\\
 Cambridge, MA 02138 \\
 \texttt{amirshanehsazzadeh@college.harvard.edu} \\
 \AND
 David Belanger \\
 Google Research \\
 \texttt{dbelanger@google.com}
 \And
 David Dohan \\
 Google Research \\
 \texttt{ddohan@google.com}
}

\begin{document}

\maketitle

\begin{abstract}
  Recently, there has been great interest in learning how to best represent proteins, specifically with fixed-length embeddings. Deep learning has become a popular tool for protein representation learning as a model's hidden layers produce potentially useful vector embeddings. TAPE~\cite{tape2019} introduced a number of benchmark tasks and showed that semi-supervised learning, via pretraining language models on a large protein corpus, improved performance on downstream tasks. Two of the tasks (fluorescence prediction and stability prediction) involve learning fitness landscapes. In this paper, we show that CNN models trained solely using supervised learning both compete with and sometimes outperform the best models from TAPE that leverage expensive pretraining on large protein datasets. These CNN models are sufficiently simple and small that they can be trained using a Google Colab notebook\footnote{ \href{https://github.com/googleinterns/protein-embedding-retrieval/blob/master/cnn_protein_landscapes.ipynb}{https://github.com/googleinterns/protein-embedding-retrieval/blob/master/cnn\_protein\_landscapes.ipynb}}. We also find for the fluorescence task that linear regression outperforms our models and the TAPE models. The benchmarking tasks proposed by TAPE are excellent measures of a model's ability to predict protein function and should be used going forward. However, we believe it is important to add baselines from simple models to put the performance of the semi-supervised models that have been reported so far into perspective. 
  
\end{abstract}

\section{Introduction}
Recent work has advocated for building protein models using semi-supervised learning~\cite{unirep, bepler, Biswas2020, tape2019, bertology, Rives2019, Lu2020}. An important question regarding these models is the effect of transfer learning: do semi-supervised models, models pretrained on a large corpus of unlabeled protein sequences, outperform supervised models? The aforementioned works all point to the affirmative. In particular, TAPE~\cite{tape2019} benchmarks a number of language models (attentive, recurrent, convolutional, UniRep multiplicative LSTM~\cite{unirep}, Bepler bidirectional LSTM~\cite{bepler}) on 5 downstream protein prediction tasks: secondary structure, contacts, remote homology, fluorescence landscape, and stability landscape. They show that pretraining a language model on a large protein corpus, specifically the Pfam database~\cite{ElGebali2018, Finn2013}, provides a significant improvement of the model on their proposed downstream tasks. Their downstream value prediction architecture, used for modeling the fluorescence and stability landscapes, learns fixed-length embeddings via an attention-weighted mean-pooling. 

Motivated by the success of semi-supervised models, we asked if purely supervised models would be effective for modeling protein fitness landscapes. We benchmark relatively small CNN models on two of the TAPE tasks: fluorescence landscape prediction and stability landscape prediction. See Table~\ref{intro} for a summary of our results and existing baselines. We find that relatively shallow CNN encoders (1-layer for fluorescence, 3-layer for stability) can compete with and even outperform the models benchmarked in TAPE. For the fluorescence task, in particular, a simple linear regression model trained on full one-hot encodings outperforms our models and the TAPE models. Additionally, 2-layer CNN models offer competitive performance with Rives et al.'s ESM (evolutionary scale modeling) transformer models~\cite{Rives2019} on $\beta$-lactamase variant activity prediction. While TAPE's benchmarking argued that pretraining improves the performance of language models on downstream landscape prediction tasks, our results show that small supervised models can, in a fraction of the time and compute required for semi-supervised models, achieve competitive performance on the same tasks.

\begin{table}[!h]
 \caption{Best model performances (Spearman $\rho$) on test sets for downstream tasks: fluorescence prediction, stability prediction, $\beta$-lactamase activity prediction. Supervised models are competitive with semi-supervised models. Note the strong performance of linear regression on one-hot encodings for the fluorescence task.}
 \label{intro}
 \centering
 \begin{tabular}{cccc}
    \toprule
    & Fluorescence & Stability & $\beta$-lactamase\\
    \hline
    \hline
    Linear Regression & \textbf{0.69} & 0.49 & 0.70\\
    \cmidrule{2-4}
    CNN & \textbf{0.69} & \textbf{0.79} & 0.87\\
    \cmidrule{2-4}
    TAPE (Non-pretrained) & 0.22 & 0.61 & -\\
    TAPE (Pretrained) & \textbf{0.68} & 0.73 & -\\
    \cmidrule{2-4}
    CPCProt (Best) & \textbf{0.68} & 0.68 & -\\
    \cmidrule{2-4}
    ESM (Best) & \textbf{0.68} & 0.71 & \textbf{0.89}\\
    ESM (Pretrained) & - & - & 0.80\\
    ESM (Fine-tuned) & - & - & \textbf{0.89}\\
    \bottomrule
 \end{tabular}
\end{table}

\section{Background}
\subsection{Proteins}
We consider proteins using only their primary structure, that is their amino acid sequence, with a 21-letter alphabet that includes the 20 standard amino acids~\cite{1984} as well as a pad index. A length $\ell$ protein $a = a_1a_2\cdots a_\ell$ is thus modeled as a discrete sequence $x = (x_1, x_2, ..., x_\ell)$ with $x_i \in \{0, 1, ..., 19\}$ and potentially padded to $(x_1, x_2, ..., x_{\ell}, 20, 20, ..., 20)$.

\subsection{Model Embeddings}
In general, a model $\Phi$ (one-hot encoding, CNN, RNN, transformer, ...) maps the encoded protein $x$ of length $\ell$ to a sequence-length-dependent array representation: $\Phi(x) \in \mathbb R^{n \times \ell}$ for some constant $n$. For learning fixed-length embeddings, the goal is to learn a mapping $\Psi$ such that $z=\Psi(\Phi(x)) \in \mathbb R^m$ with $m$ being independent of $\ell$. We also want $z$ to be a ``useful'' protein embedding, in the sense of having strong signal for some downstream feature(s).

Our approach to computing and learning embeddings from a model $\Phi$ is to apply a potentially learnable mapping from the encoded representations, which are arrays of the shape $(n, \ell)$, to fixed-length arrays of the shape $(m,)$. The simplest mapping we consider is pooling, which is non-learnable and involves either taking the average (MeanPool) or the maximum (MaxPool) over the length-dependent axis of $\Phi(x) \in \mathbb R^{n\times\ell}$. A learnable, but simple, mapping involves a single dense layer with weight matrix $W \in \mathbb R^{m \times n}$ and bias vector $b \in \mathbb R^m$ and a non-linear activation $\phi$. Then we consider the array of transformed amino acid level embeddings $$\left[\phi(W\Phi(x)^{(i)}+b)\right]_{i=1}^{\ell} \in \mathbb R^{m \times \ell}$$ and apply a pooling operation. If we apply MaxPool we call this operation LinearMaxPool and likewise for MeanPool. Note that we only use the ReLU activation for $\phi$. At a high-level our approach is motivated by a similar technique from NLP known as contextual lenses~\cite{lenses}.

\subsection{Protein Landscape Prediction Tasks}
Two of the TAPE tasks, fluorescence prediction and stability prediction, are scalar value prediction problems~\cite{tape2019}. Their prediction model involves taking the sequence-length-dependent array representation outputted by a language model, applying an attention-weighted mean-pooling operation to get a sequence-length-independent vector representation, applying a dense hidden layer of size 512 followed by ReLU activation, and finally performing value prediction using a dense output layer. The task-specific learning produces fixed-length embeddings designed to be classified by scalar value (fluorescence or stability) using a small MLP trained with mean squared error (MSE) loss.

We also consider a variant prediction task from Rives et al~\cite{Rives2019}. The publicly available implementation\footnote{\href{https://github.com/facebookresearch/esm/blob/master/examples/variant_prediction.ipynb}{https://github.com/facebookresearch/esm/blob/master/examples/variant\_prediction.ipynb}} uses precomputed pretrained transformer protein embeddings and their paper includes the results from a transformer model that is both pretrained and fine-tuned.

It is worth noting that the tasks we consider are fine-grained landscape prediction, where we attempt to predict the output of a real-valued functional assay (e.g. brightness). Works such as Biswas et. al.~\cite{Biswas2020} consider more coarse-grained tasks where the goal is to predict the presence or lack of function.

We give a brief overview of the prediction tasks and the datasets used. For more detail, particularly from the biological perspective, refer to~\cite{sarkisyan2016, rocklin2017, Gray2018}.

\subsubsection{Fluorescence Landscape Prediction}
This regression task involves mapping a protein to its log-fluorescence, which is a real-valued label. The experimental data is from Sarkisyan et al.~\cite{sarkisyan2016} and the curated dataset is from TAPE\footnote{\href{https://github.com/songlab-cal/tape}{https://github.com/songlab-cal/tape}}. The data consists of mutated variants of a wild-type GFP protein with edit distance up to 14. The train set contains all variants within edit distance 3 of the wildtype (at most 3 mutations away) and the test set contains all variants at least 4 mutations away from the wildtype. This split by edit distance allows for testing the generalizability of a model trained on a small (local) neighborhood of the wildtype to a larger (global) neighborhood. Note that because of this split the train set consists of 82\% bright proteins (log-fluorescence greater than 2.5) and 18\% dark proteins (log-fluorescence less than 2.5) whereas the test set consists of only 32\% bright proteins and 68\% dark proteins. This class imbalance makes it difficult for models to generalize from the low mutation train data to the high mutation test data, as evidenced by the TAPE results and our results. The primary metric of interest is Spearman's rank correlation coefficient $\rho$ on the test set. MSE on the test set is also reported. Additionally, due to the class imbalance $\rho$ and MSE on the bright and dark subsets of the test set are reported.

\subsubsection{Stability Landscape Prediction}
This regression task involves mapping a protein to its stability, a real-valued label, which is a measurement of the log-concentration of protease required for $\nicefrac{1}{2}$ of the cells cultured with said protein to satisfy a collection threshold, normalized against the corresponding predicted log-concentration. The experimental data is from Rocklin et al.~\cite{rocklin2017} and the curated dataset is from TAPE. The train set contains proteins from experimental design rounds whereas the test set contains proteins 1 mutation away from top candidates. This task thus tests for the ability of a model to learn local properties from global information, the opposite of the fluorescence task. The primary metric of interest is again Spearman's rank correlation coefficient $\rho$ on the test set. Additionally, an accuracy metric is reported. This accuracy measures the degree to which binarized model predictions agree with binarized true values. Specifically, a prediction is said to be accurate in 2 of 4 cases:
\begin{enumerate}[label=(\roman*)]
  \item The variant protein stability is greater than the parent protein stability and the predicted variant stability is greater than the predicted parent stability.
  \item The variant protein stability is less than the parent protein stability and the predicted variant stability is less than the predicted parent stability.
\end{enumerate}
The proteins primarily consist of 4 fold topologies: $\{\alpha\alpha\alpha, \alpha\beta\beta\alpha, \beta\alpha\beta\beta, \beta\beta\alpha\beta\beta\}$ and so $\rho$ and the accuracy metric are reported on the full test set as well as the test set restricted to each of these topologies.

\subsubsection{$\beta$-lactamase Variant Prediction}
The data for this task comes from Envision (Gray et. al.)~\cite{Gray2018} and is curated by Rives et al. It consists of mutated $\beta$-lactamase sequences as inputs and scaled mutation effects as targets. We report mean and standard deviation of Spearman's $\rho$ on a partition of size 5 of the generated test set. The train set is also varied and generated by random splits of 1\%, 10\%, 30\%, 50\%, and 80\% of the data. FAIR's publicly available implementation measures the performance of pretrained, but not fine-tuned, transformer embeddings with an 80-20 data split. They measure performance on the whole test set. To compare with their full paper we adapt their implementation by using train sets ranging from 1\% to 80\% of the data and create a partition of size 5 of the test set. 

\section{Related Work}
The most directly related work is TAPE~\cite{tape2019}. which curates the datasets for and formulates the specific fluorescence and stability landscape prediction problems we consider. The experimental data for the fluorescence task is from Sarkisyan et al.~\cite{sarkisyan2016}. They model the GFP landscape using linear regression on single mutation effects, multiple regression which is similar to their linear regression but models fluorescence as a sigmoidal function of mutations, and MLPs where the inputs are binarized according to the presence of mutations in the protein genotype. The experimental data for the stability task comes from Rocklin et al.~\cite{rocklin2017}. Their library design is computational, but uses energy-based and molecular dynamics approaches as opposed to primary sequence modeling.

Lu et al.~\cite{Lu2020} develop CPCProt, an autoregressive model pretrained on the same protein corpus as TAPE using mutual information maximization. Their model is substantially smaller ($\sim$1.7M parameters) than the TAPE models, which range from $\sim$19M-182M parameters, but its self-supervised embeddings provide comparable downstream performance using TAPE's attention-weighted mean architecture.

Rives et al.~\cite{Rives2019} does similar work but trains varying size transformer language models on 250 million UniRef~\cite{Suzek2014} sequences. Their transformer embeddings achieve state-of-the-art performance on a number of downstream tasks such as contact, secondary structure, and remote homology. They also show that larger model size, which results in better protein language modeling performance, improves downstream performance. We consider their variant prediction task, specifically for $\beta$-lactamase data from Envision~\cite{Gray2018}.

We are unaware of other directly comparable works that use the same datasets, but machine learning techniques have been used in a similar fashion to learn useful protein embeddings in~\cite{Yang2018, unirep, bepler, bertology, Rives2019, bertology}.

\section{Models and Training Procedure}
We use linear regression, specifically scikit-learn's~\cite{scikit-learn} ridge regression, on the full one-hot protein encoding as a baseline for all tasks. It is important to note that the one-hot baseline in TAPE~\cite{tape2019} does not use the full one-hot encoding but instead uses a bag-of-words representation (MeanPool applied to the full one-hot encoding), which measures the frequency of amino acids. We refer to this as One-hot (AA counts). This pooling results in a substantial loss of information, especially for the fluorescence task which is trained on proteins that are very close in edit distance.

Our supervised models only rely on 1-D convolution layers, dense layers, and ReLU activations. Training is done using the Adam optimizer~\cite{adam} with variable learning rates and weight decays per architecture component, but with no learning rate warmup. Mean squared error loss is used. Sequences are padded to maximum length and the corresponding padded components are zeroed out before pooling.

\subsection{Fluorescence Models}
For modeling the fluorescence landscape we use a 1-layer 1-D CNN with feature size 1024 and kernel size 5 as an encoder. ReLU is applied after the convolution layer. To create embeddings we use either MaxPool (embedding dimension 1024) or LinearMaxPool with an embedding dimension of 2048. For the predictor a dense layer is used. Our fluorescence CNN models have $\sim$110k-220k parameters, which is $\sim$1 order of magnitude less than CPCProt~\cite{Lu2020} and $\sim$2-3 orders of magnitude less than the TAPE models. We also use a linear regression model with $\sim$5k parameters as a baseline. We report the hyperparameters in Table~\ref{model_params}. Note that Ep is epochs, BS is batch size, Dim is embedding dimension, and the learning rates and weight decays are reported as tuples with the values corresponding to the encoder, mapping, and predictor in that order.

\subsection{Stability Models}
For modeling the stability landscape we use a 3-layer 1-D CNN with feature sizes all 1024 and kernel sizes all 5 as an encoder. ReLU is applied between and after convolution layers. To create embeddings we use either MaxPool (embedding dimension 1024) or LinearMaxPool with an embedding dimension of 2048. We also try a Dilated CNN + MaxPool model with kernel dilation of 2 in the 2nd layer. For the predictor a dense layer is used. Our stability CNN models have $\sim$10.6M-12.7M parameters, which is $\sim$1 order of magnitude greater than CPCProt and $\sim$0-1 orders of magnitude less than the TAPE models. Additionally, we ensemble the above three models by averaging their predictions to create an ``Ensemble of CNNs'' model. The motivation behind this averaging is the variance in individual model performance across different fold toplogies. We also use a linear regression model with $\sim$1k parameters as a baseline. We report the hyperparameters in Table~\ref{model_params}.

\subsection{$\beta$-lactamase Variant Models}
For modeling the $\beta$-lactamase variant landscape we use a 2-layer 1-D CNN with feature sizes 1024 and 512 (layer 1 and layer 2) and kernel sizes 9 and 7 (layer 1 and layeer 2) as an encoder. ReLU is applied between and after convolution layers. To create embeddings we use MaxPool (embedding dimension 512). We also try a Dilated CNN + MaxPool model with kernel dilation of 2 in the 1st layer (we at times refer to this as ``D 2-Layer + MaxPool''). For the predictor a dense layer is used. Our variant CNN models have $\sim$38.7M parameters. Additionally, we ensemble the above two models by averaging their predictions to create an Ensemble of CNNs model. We also use a linear regression model with $\sim$6k parameters as a baseline. We report the hyperparameters in Table~\ref{model_params}.

\setlength{\tabcolsep}{3pt}
\begin{table}[!h]
 \caption{Model hyperparameters}
 \label{model_params}
 \centering
 \begin{tabular}{ccccccc}
    \toprule
    Task & Model & Ep & BS & Dim & Learning Rates & Weight Decays\\
    \hline
    \hline
    \multirow{2}{*}{GFP} & 1-Layer CNN + MaxPool & 50 & 256 & 1024 & (1e-3, 0, 5e-6) & (0, 0, 0.05)\\
     & 1-Layer CNN + LinearMaxPool & 50 & 256 & 2048 & (1e-3, 5e-5, 5e-6) & (0, 0.05, 0.05)\\
    \cmidrule{2-7}
    \multirow{3}{*}{Stability} & 3-Layer CNN + MaxPool & 5 & 16 & 1024 & (5e-4, 0, 5e-5) & (0.025, 0, 0.025)\\
     & 3-Layer Dilated CNN + MaxPool & 5 & 16 & 1024 & (5e-4, 0, 5e-5) & (0.025, 0, 0.025)\\
     & 3-Layer CNN + LinearMaxPool & 10 & 256 & 2048 & (1e-5, 5e-5, 5e-6) & (0, 0, 0)\\
     \cmidrule{2-7}
     \multirow{2}{*}{Variant} & 2-Layer CNN + MaxPool & 500 & 32 & 512 & (1e-3, 0, 1e-3) & (0.1, 0, 0.05)\\
     & 2-Layer Dilated CNN + MaxPool & 500 & 32 & 512 & (1e-3, 0, 1e-3) & (0.1, 0, 0.05)\\
    \bottomrule
 \end{tabular}
\end{table}

\section{Results}
For the TAPE landscape prediction tasks, we report results in a similar fashion to TAPE~\cite{tape2019} by recreating an analog of their tables with our results and the best of their results over all models. We also include the results provided in CPCProt~\cite{Lu2020}. For the $\beta$-lactamase prediction task, we recreate an analog of Table S7 in Rives et al.~\cite{Rives2019}.

\subsection{Fluorescence Results}
Table~\ref{gfp_results} contains mean squared error (MSE) and Spearman's $\rho$ for the full GFP test set as well as the bright subset (log-fluorescence greater than 2.5) and dark subset (log-fluorescence less than 2.5). For the TAPE results we present the best reported metric across a class of models. Note the significant difference between our one-hot linear regression baseline and the one-hot baseline of TAPE. This is because we learn a regressor on the full one-hot vector whereas TAPE uses a bag-of-words representation, which is the full one-hot vector with MeanPool applied. For the CPCProt results, we provide the best reported metric.

\setlength{\tabcolsep}{6pt}
\begin{table}[!h]
 \caption{Fluorescence prediction results. $\rho$ is Spearman's rank correlation coefficient.}
 \label{gfp_results}
 \centering
 \begin{tabular}{cccccccc}
  \toprule
   & & \multicolumn{2}{c}{Full Test Set} & \multicolumn{2}{c}{Bright Mode Only} & \multicolumn{2}{c}{Dark Mode Only}\\
   \cmidrule(lr){3-4} \cmidrule(lr){5-6} \cmidrule(lr){7-8}
   & & MSE & $\rho$ & MSE & $\rho$ & MSE & $\rho$\\
   \hline
   \hline
   Baseline & Linear Regression & 0.35 & \textbf{0.69} & 0.09 & \textbf{0.68} & 0.33 & \textbf{0.05}\\
   \cmidrule{2-8}
   \multirow{2}{*}{CNN} & 1-Layer + MaxPool & 0.26 & \textbf{0.69} & 0.09 & 0.65 & 0.29 & \textbf{0.05} \\
    & 1-Layer + LinearMaxPool & 0.23 & \textbf{0.69} & 0.12 & 0.66 & 0.28 & \textbf{0.05} \\
   \cmidrule{2-8}
   \multirow{4}{*}{TAPE} & One-hot (AA counts) & 2.69 & 0.14 & 0.08 & 0.03 & 3.95 & 0.0\\
    & Non-pretrained models & 2.35 & 0.22 & \textbf{0.07} & 0.08 & 3.43 & 0.0\\
    & Pretrained models & \textbf{0.19} & \textbf{0.68} & 0.08 & 0.63 & \textbf{0.22} & \textbf{0.05}\\
    & Best of all models & \textbf{0.19} & \textbf{0.68} & \textbf{0.07} & 0.63 & \textbf{0.22} & \textbf{0.05}\\
    \cmidrule{2-8}
    CPCProt & Best of all models & 0.81 &\textbf{0.68} & - & - & - & -\\
    \cmidrule{2-8}
    ESM & Best of all models & - &\textbf{0.68} & - & - & - & -\\
  \bottomrule
 \end{tabular}
\end{table}

\begin{figure}[!h]
 \centering
 \includegraphics[width=0.45\linewidth]{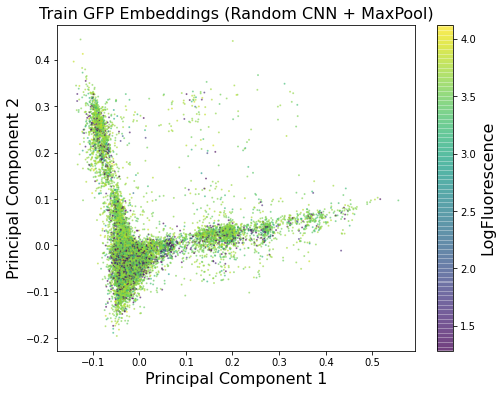}
 \hspace{4.5mm}
 \includegraphics[width=0.45\linewidth]{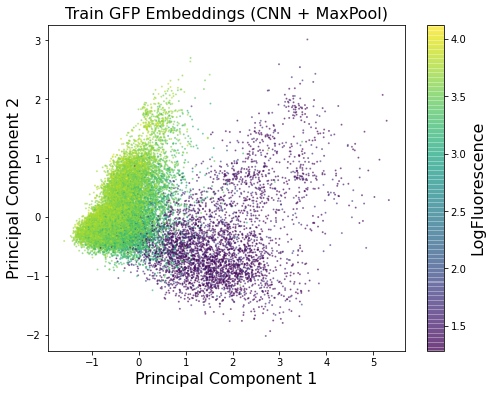}
 \\
 \vspace{3.5mm}
 \includegraphics[width=0.45\linewidth]{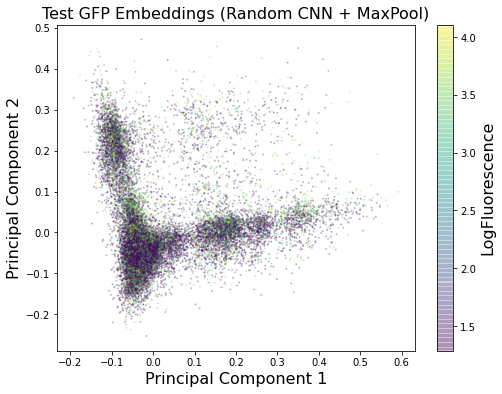}
 \hspace{4.5mm}
 \includegraphics[width=0.45\linewidth]{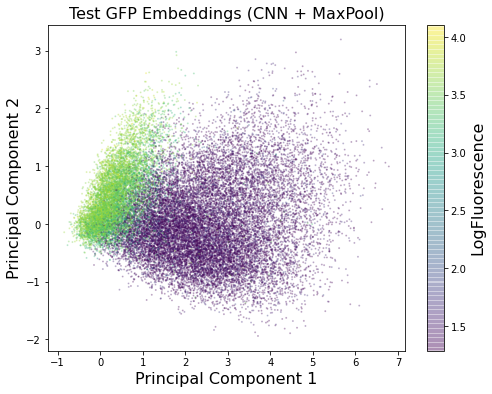}
 \caption{PCA of GFP train and test set embeddings colored by log-fluorescence for randomly initialized CNN model and trained CNN model. Supervised training produces embeddings that separate both the train and test sets according to log-fluorescence.}
 \label{gfp_pca_plot}
\end{figure}

Figure~\ref{gfp_pca_plot} shows the top 2 principal components of the model embeddings colored by log-fluorescence. We separate the sequences by the train and test split and fit the PCA only to the train set. For models we use a trained CNN + MaxPool model and an identical, but randomly initialized, CNN + MaxPool model. Notice that the trained model separates sequences by log-fluorescence whereas the randomly initialized model does not.

We see from Table~\ref{gfp_results} that the CNN architectures outperform the TAPE models and CPCProt~\cite{Lu2020} in the more relevant Spearman $\rho$ metric. As shown in Figure~\ref{gfp_pca_plot}, the 1-layer CNN models are able to learn fixed-length embeddings that separate the proteins based on log-fluorescence. It is worth noting that due to the bimodality of the data, MSE is not a particularly good metric. For example, the TAPE model that achieves the lowest bright mode MSE of 0.07 has predictions that are negatively rank-correlated with the true fluorescence values. Furthermore, a simple linear regression model outperforms every other reported model.

\subsection{Stability Results}
Table~\ref{stab_results} shows overall performance on the stability test set, and Tabel~\ref{stab_top_results} shows performance on specific fold topologies. In addition to reporting Spearman's $\rho$, we report the previously defined accuracy measurement that uses the parent protein as a decision boundary and labels mutations as beneficial if predicted stability is greater than predicted parent stability and deleterious if the opposite is true. We again present the best reported metric across a class of models for the TAPE results and the best reported metric for CPCProt.

\begin{table}[!h]
 \caption{Overall stability prediction results}
 \label{stab_results}
 \centering
 \begin{tabular}{cccc}
  \toprule
    & & Spearman's $\rho$ & Accuracy\\
    \hline
    \hline
    Baseline & Linear Regression & 0.49 & 0.60\\
    \cmidrule{2-4}
    \multirow{4}{*}{CNN} & 3-Layer + MaxPool & 0.76 & 0.75 \\
     & Dilated 3-Layer + MaxPool & 0.75 & 0.73 \\
    & 3-Layer + LinearMaxPool & 0.71 & \textbf{0.77} \\
     & Ensemble of CNNs & \textbf{0.79} & \textbf{0.77}\\
    \cmidrule{2-4}
    \multirow{4}{*}{TAPE} & One-hot (AA counts) & 0.19 & 0.58\\
    & Non-pretrained models & 0.61 & 0.68\\
    & Pretrained models & 0.73 & 0.70\\
    & Best of all models & 0.73 & 0.70\\
    \cmidrule{2-4}
    CPCProt & Best of all models & 0.68 & -\\
    \cmidrule{2-4}
    ESM & Best of all models & 0.71 & -\\
  \bottomrule
 \end{tabular}
\end{table}

\setlength{\tabcolsep}{5.25pt}
\begin{table}[!h]
 \caption{Stability prediction results broken down by protein topology}
 \label{stab_top_results}
 \centering
 \begin{tabular}{cccccccccccc}
  \toprule
    & & \multicolumn{2}{c}{$\alpha\alpha\alpha$} & \multicolumn{2}{c}{$\alpha\beta\beta\alpha$} & \multicolumn{2}{c}{$\beta\alpha\beta\beta$} & \multicolumn{2}{c}{$\beta\beta\alpha\beta\beta$}\\
    \cmidrule(lr){3-4} \cmidrule(lr){5-6} \cmidrule(lr){7-8} \cmidrule(lr){9-10}
    & & $\rho$ & Acc & $\rho$ & Acc & $\rho$ & Acc & $\rho$ & Acc\\
    \hline
    \hline
    Baseline & Linear Regression & 0.21 & 0.66 & -0.03 & 0.6 & 0.51 & 0.64 & 0.38 & 0.61\\
    \cmidrule{2-10}
    \multirow{4}{*}{CNN} & 3-Layer + MaxPool & 0.69 & \textbf{0.71} & 0.37 & 0.70 & 0.50 & 0.72 & 0.60 & 0.68\\
    & Dilated 3-Layer + MaxPool & 0.67 & 0.69 & 0.49 & 0.69 & 0.61 & 0.70 & 0.53 & 0.64\\
    & 3-Layer + LinearMaxPool & 0.59 & 0.69 & 0.52 & 0.77 & 0.55 & 0.73 &0.6 & \textbf{0.70}\\
     & Ensemble of CNNs & 0.67 & \textbf{0.71} & \textbf{0.53} & 0.75 & 0.65 & \textbf{0.74} & 0.60 & \textbf{0.70}\\
    \cmidrule{2-10}
    \multirow{4}{*}{TAPE} & One-hot (AA counts) & 0.58 & 0.59 & 0.04 & 0.58 & -0.05 & 0.58 & 0.54 & 0.58\\
    & Non-pretrained models & 0.64 & 0.69 & 0.39 & 0.70 & 0.63 & 0.67 & 0.65 & 0.67\\
    & Pretrained models & \textbf{0.72} & 0.70 & 0.48 & \textbf{0.79} & \textbf{0.68} & 0.71 & \textbf{0.67} & \textbf{0.70}\\
    & Best of all models & \textbf{0.72} & 0.70 & 0.48 & \textbf{0.79} & \textbf{0.68} & 0.71 & \textbf{0.67} & \textbf{0.70}\\
  \bottomrule
 \end{tabular}
\end{table}

\begin{figure}[!h]
 \centering
 \includegraphics[width=0.45\linewidth]{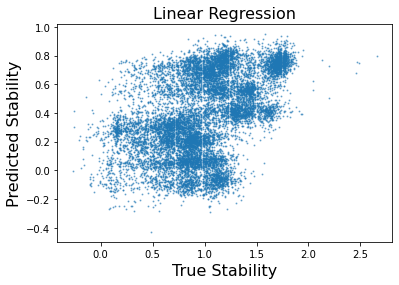}
 \hspace{4mm}
 \includegraphics[width=0.45\linewidth]{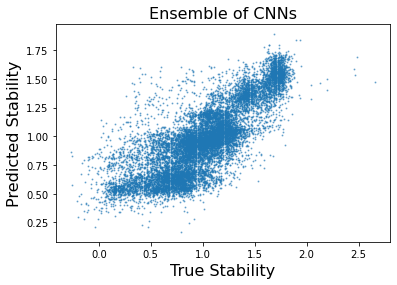}
 \caption{Predicted vs. true stabilities for linear regression model and ensemble of CNNs.}
 \label{stab_plot}
\end{figure}

Figure~\ref{stab_plot} shows plots of the predicted stabilities from the linear regression model and from the Ensemble of CNNs.
 
We see from Table~\ref{stab_results} that the CNN ensemble model outperforms the TAPE models and CPCProt on the full dataset. Table~\ref{stab_top_results} shows that our 3-layer CNN models are all competitive with the TAPE models and in fact outperform on 3 out of 8 metrics. In particular, our models achieve better binarized accuracies indicating that they are able to effectively predict whether or not a mutation will be beneficial or deleterious. This is perhaps more relevant than rank-correlation, since in general for protein engineering the aim is to predict whether or not a mutation will be beneficial or deleterious as opposed to \emph{how} beneficial or deleterious said mutation will be. Interestingly, for this task, linear regression performs quite poorly. This is likely because the task requires the model to learn from broader, more global data and localize information which is the opposite of what the fluorescence task requires.

\subsection{$\beta$-lactamase Variant Results}
Table~\ref{beta_lac_results} compares the performance of our CNN models with the ESM transformer models~\cite{Rives2019}. We report mean and standard deviation of Spearman $\rho$ across a size 5 partition of the test sets and present results for the fine-tuned ESM model from the corresponding paper.

\setlength{\tabcolsep}{2.55pt}
\begin{table}[!h]
 \caption{Comparison across models of mean and standard deviation $\beta$-lactamase variant prediction performance, measured by Spearman $\rho$, on a partition of size 5 of the test set.}
 \label{beta_lac_results}
 \centering
 \begin{tabular}{ccccccc}
  \toprule
   & & 1\% data & 10\% data & 30\% data & 50\% data & 80\% data\\
   \hline
   \hline
   Baseline & Linear Regression & $0.19 \pm 0.10$ & $0.52 \pm 0.01$ & $0.67 \pm 0.02$ & $0.70 \pm 0.02$ & $0.70 \pm 0.03$\\
   \cmidrule{2-7}
   \multirow{3}{*}{CNN} & 2-Layer + MaxPool & $0.24 \pm 0.03$ & $0.59 \pm 0.02$ & $0.77 \pm 0.01$ & $0.83 \pm 0.02$ & $0.86 \pm 0.01$\\
   & D 2-Layer + MaxPool & $0.15 \pm 0.03$ & $0.62 \pm 0.02$ & $0.76 \pm 0.02$ & $0.83 \pm 0.02$ & $0.86 \pm 0.01$\\
    & Ensemble of CNNs & $0.23 \pm 0.03$ & $0.62 \pm 0.02$ & $0.78 \pm 0.01$ & $0.84 \pm 0.02$ & $0.87 \pm 0.01$\\
   \cmidrule{2-7}
   \multirow{2}{*}{ESM} & Pretrained & $0.51 \pm 0.03$ & $0.68 \pm 0.02$ & $0.75 \pm 0.01$ & $0.78 \pm 0.01$ & $0.80 \pm 0.02$\\
   & Fine-tuned & $0.39 \pm 0.03$ & $0.69 \pm 0.01$ & $0.84 \pm 0.01$ & $0.88 \pm 0.01$ & $0.89 \pm 0.01$\\
  \bottomrule
 \end{tabular}
\end{table}

\begin{figure}[!h]
 \centering
 \includegraphics[width=0.45\linewidth]{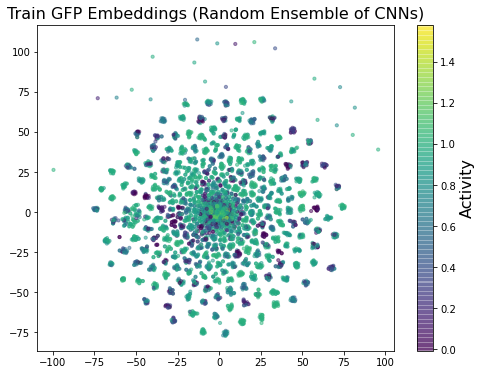}
 \hspace{4.5mm}
 \includegraphics[width=0.45\linewidth]{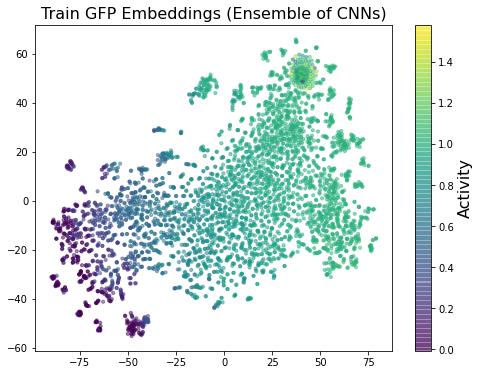}
 \\
 \vspace{3.5mm}
 \includegraphics[width=0.45\linewidth]{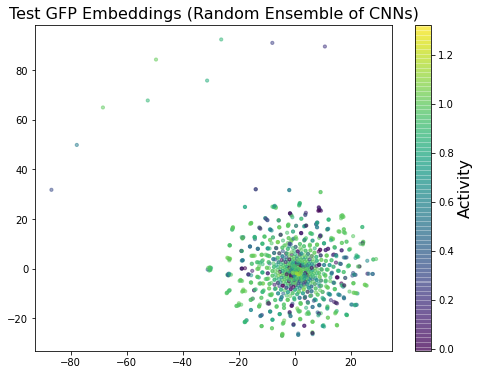}
 \hspace{4.5mm}
 \includegraphics[width=0.45\linewidth]{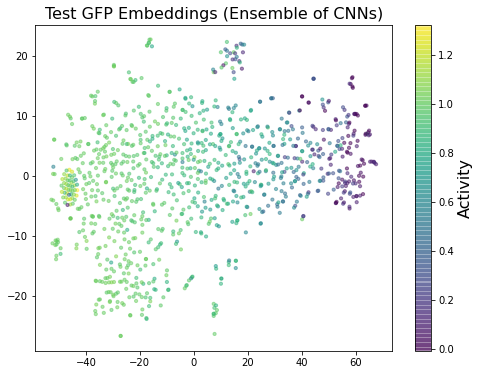}
 \caption{t-SNE of $\beta$-lactamase train and test set embeddings colored by activity for randomly initialized ensemble of CNN models and trained ensemble of CNN models. Supervised training produces embeddings that separate both the train and test sets according to activity.}
 \label{variant_tsne_plot}
\end{figure}

Figure~\ref{variant_tsne_plot} shows a 2-D t-SNE of the model embeddings colored by activity. We separate the sequences by the 80-20 train and test split. For models, we use a trained ensemble of CNN + MaxPool models and an identical, but randomly initialized, ensemble of CNN + MaxPool models. Again, notice that the trained ensemble of models separates sequences by activity whereas the randomly initialized ensemble does not.

We see in Table~\ref{beta_lac_results} that a 2-layer CNN with an 80-20 train test split outperforms the pretrained ESM transformer~\cite{Rives2019} and is competitive with the pretrained and fine-tuned ESM transformer. The t-SNE plots in Figure~\ref{variant_tsne_plot} shows that the Ensemble of CNNs embeddings, when trained, separate the variants by actiivity. All deep models outperform linear regression by a noticeable margin. It is worth noting, however, that with lower percentages of the data used for training, the ESM models considerably outperform our CNN models. Pretraining the ESM transformer resulted in embeddings that split the data modestly by activity without any downstream supervision, which we believe explains this advantage with less data.

\section{Discussion}
We see that relatively simple and small CNN models trained entirely with supervised learning for fluorescence or stability prediction compete with and outperform the semi-supervised models benchmarked in TAPE~\cite{tape2019}, despite requiring substantially less time and compute. While the benefit of transfer learning via pretraining is evident, the performance of the pretrained language models does not appear to justify the cost of pretraining for these protein landscape modeling tasks. Furthermore, it appears that with a sufficient amount of data, CNNs trained only with supervised learning can compete with the pretrained (and fine-tuned) transformers from Rives et. al.~\cite{Rives2019} on $\beta$-lactamase variant prediction. However, the semi-supervised models outperform considerably on this task when training data is limited. This leads us to conjecture that pretraining may be most useful for downstream tasks with limited supervised data.

\bibliographystyle{plain}
\bibliography{main}

\end{document}